# SUB-NYQUIST CO-PRIME SENSING WITH COMPRESSED INTER-ELEMENT SPACING - LOW LATENCY APPROACH

*Abstract*

*Co-prime arrays with compressed inter-element spacing (CACIS) is one of the generalizations of the co-prime array. The inter-element spacing can be varied in this case. The prototype co-prime arrays and nested arrays are a special case of the CACIS scheme. The problems that were not addressed previously are considered in this paper. The fundamentals of the difference set for the CACIS configuration are developed for low latency. In addition, the closed-form expressions for the weight function (number of samples that contribute to estimate the autocorrelation) and bias window of the correlogram estimate, which were previously unknown, are derived. Ideally, the bias window should be an impulse. Several examples are provided along with simulations to verify the claims made. All possible sample pairs are used for estimation, which provides for low latency. As an application, temporal spectrum is considered for simulations.*

## 1. INTRODUCTION

Co-prime arrays (or samplers) utilize two sub-arrays (or sub-samplers) with sampling periods that are *M* and *N* times the Nyquist period. *M* and *N* are co-prime integers [1]. These structures cannot give back the original time-domain Nyquist signal, however, it can generate the autocorrelation at the Nyquist rate. There are many applications where the time-domain signal is not necessary, only the second order statistics (e.g. autocorrelation, cross-correlation, spectrum) is of importance. Initially, minimum redundancy arrays were proposed in [2] which attempts to minimize the redundant spacings. Later, the nested array configuration was proposed that was able to achieve $O(N^2)$ degrees of freedom with just *N* points [3]. An alternative to the nested array can be found in the form of co-prime arrays and can resolve $O(MN)$ differences (i.e. lags) in the autocorrelation function [1]. The nested array configuration required one array to have an inter-element spacing similar to the Nyquist framework and is no longer required in the case of the co-prime array configuration. Some of the applications are in the area of sub-Nyquist spectrum sensing, beam-forming, direction of arrival estimation, range and velocity estimation, etc. [4-16].

Two generalizations of the co-prime array were proposed in [17-18] through the compression of the inter-element spacing and displacement of the sub-arrays. The former is referred to as the Co-prime Array with Compressed Inter-element Spacing (CACIS) and will be discussed in this paper. The prototype co-prime array and nested arrays are special cases of the CACIS scheme. Most of the work in the past required large latency. Low latency based estimation of damped complex exponential modes was studied in [19]. Power spectrum estimation using the combined difference set with low latency was shown in [5]. It is a part of a larger narrative on low latency co-prime sensing described in [20]. It includes the difference set analysis for the ideal and perturbed prototype co-prime array. It also studies the co-prime array with multiple periods, co-prime correlogram spectral estimation, and proposes cross-correlation and cross-spectrum based applications. Low latency extended or conventional co-prime array is studied in [21]. Motivated by the recent low latency findings, this paper considers the difference set for the CACIS configuration from a low latency perspective. A summary/contribution of this paper is mentioned below:

- The basics of the sub-Nyquist co-prime scheme is discussed in Section 2. First the Nyquist structure in presented, followed by the prototype co-prime structure, and finally the CACIS structure is introduced.

- The fundamentals of the difference set for the CACIS configuration, namely, uniqueness, range, and continuity of difference values are developed in Section 3-4.

- The closed-form expression for the number of samples that contribute to the autocorrelation estimate at each difference value (i.e. the weight function) is developed.

Author: Usham V. Dias
Dept. of Electrical Engineering, Indian Institute of Technology Delhi.

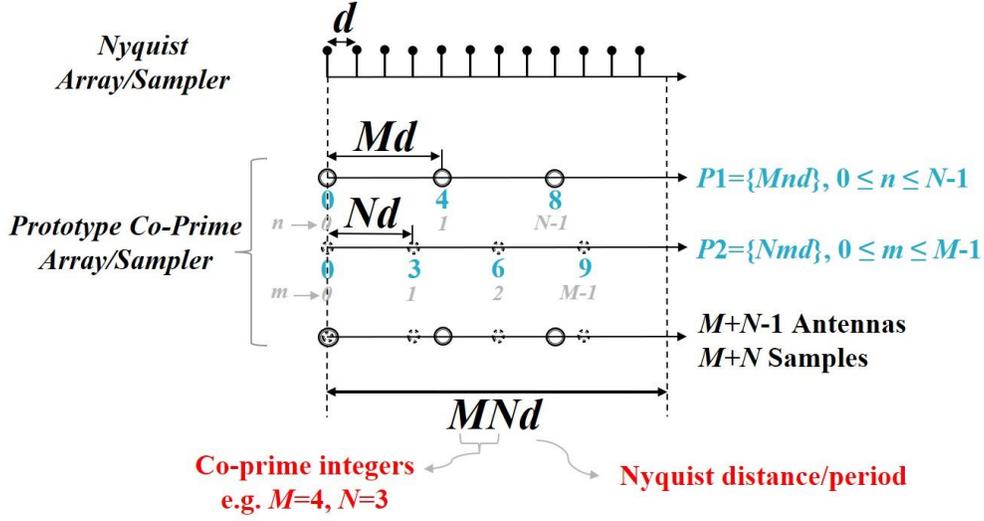

Fig. 1: Nyquist and prototype co-prime acquisition.

- The closed-form expression for the bias window of the correlogram spectral estimate is also derived.

- Examples of simulated weight functions and bias windows are provided to verify the theory developed.

- Spectral estimation is demonstrated with low latency for temporal signals.

- The paper concludes with several possible future directions and necessary references.

## 2. BASICS OF CACIS

According to the Nyquist sampling theorem, the analog input signal should be sampled at a rate that is greater than twice the highest frequency of the input signal. Therefore, $f_s > 2f_m$ where $f_s$ is the sampling frequency and $f_m$ is the highest frequency in the input signal. This implies that the distance between samples (sampling time period) is $d = 1/f_s$. Note that the sampling theorem is also valid for the spatial domain where the distance between the uniform antenna array elements is $d=\lambda/2$ (Refer Fig. 1).

Sub-Nyquist structures attempt to reduce the number of samples acquired or the number of antennas. Thus, reducing the cost for acquisition. The sub-Nyquist prototype co-prime array does not use a single uniform array but two sub-arrays which sparsely sample the signal with distance $Md$ and $Nd$. Here, $M$ and $N$ are co-prime (Refer Fig. 1). The two samplers individually are uniform (but skips several samples). Note that the combined sampling pattern is not uniform and has missing samples (sub-Nyquist). The overlapping zeroth location has only one antenna (common to both sub-arrays). Therefore, M+N-1 antennas are present. For samplers, the zeroth location has two samples (one is redundant) and are kept so as to maintain synchronism between the sub-samplers and uniform sampling patterns independently. Therefore, M+N samples are acquired (one redundant).

Despite the missing samples the autocorrelation of the acquired sub-Nyquist signal can generate most of the difference values or lags. This implies that the second order statistics likes autocorrelation, cross-correlation, spectrum, etc. can be estimated. Note that the Fourier transform of the autocorrelation gives the correlogram spectral estimate (details can be found in [20]). Sub-Nyquist co-prime arrays and samplers with compressed inter-element spacing (CACIS) is one of the generalizations of the prototype co-prime array and is considered in this paper.

The prototype co-prime array seems to be well studied in the literature. The focus in this paper will be on the CACIS configuration of co-prime sensing [17]. The spacing between the elements of one sub-array can be varied, while keeping the other sub-array fixed. This framework uses a co-prime pair ($\breve{M}$, $N$). Note that there is a relationship between $M$ (used in the prototype co-prime scheme) and $\breve{M}$, i.e. $M$ is a product of two positive integers $\breve{M}$ and $p$, where $p$ is the integer compression factor such that $p \in [1, M]$:

$$M = p\breve{M} \qquad \text{and} \qquad \breve{M} = \frac{M}{p} \qquad (1)$$

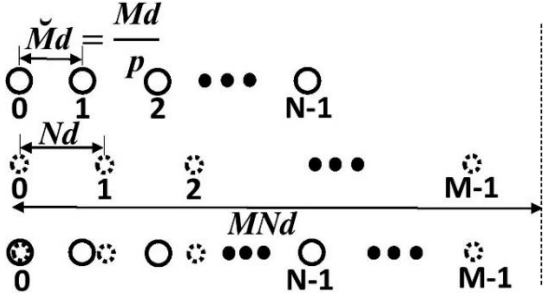

Fig. 2: Sub-Nyquist CACIS acquisition.

Thus, $p=1$ gives the prototype co-prime array configuration. Compression in the inter-element spacing is obtained for $2 \leq p \leq M$ with $p = M$ resulting in the nested array configuration. Since $M$ and $N$ are selected to be co-prime, $\breve{M}$ and $N$ are also co-prime. Though trivial, it may be noted that $\breve{M} \leq M$. In our discussion we assume that the $N$-element sub-array is compressed while the $M$-element sub-array is uncompressed. The compressed sub-array has an inter-element spacing of $\breve{M}d=Md/p$ and their locations are given by $\breve{M}nd$ where $0 \leq n \leq N$-$1$ as in Fig. 2. From an antenna array perspective, the total number of antenna elements or sensors in this scheme is $M+N$-$1$, excluding the overlapping zeroth element. But, from a sampling perspective the number of samples in a single co-prime period will be $M+N$ with the first sample at the zeroth location being retained for synchronization.

## 3. CACIS DIFFERENCE SET

Let $x(\breve{M}n)$ and $x(Nm)$ be the acquired data from the two uniform co-prime sub-arrays (or sub-samplers) where $M=p\breve{M}$. The difference set tells us about the autocorrelation lags (referred to as difference value $l$). Some of the questions are: can all the lags be generated? Are there missing lag values? What is the continuous range of values generated? How many unique pairs are available to generate each lag value? etc. Self-differences are the difference values or lags generated by the individual sub-arrays or sub-samplers. The set of self-differences generated by each of these samplers denoted by $\mathcal{L}^+_{S\breve{M}}$ and $\mathcal{L}^+_{SN}$, are given below:

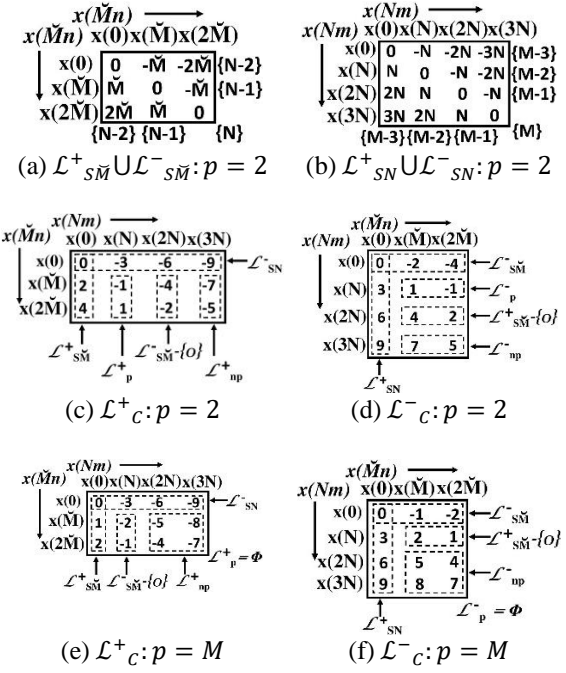

(a) $\mathcal{L}^+_{S\breve{M}} \cup \mathcal{L}^-_{S\breve{M}}: p = 2$    (b) $\mathcal{L}^+_{SN} \cup \mathcal{L}^-_{SN}: p = 2$

(c) $\mathcal{L}^+_C: p = 2$    (d) $\mathcal{L}^-_C: p = 2$

(e) $\mathcal{L}^+_C: p = M$    (f) $\mathcal{L}^-_C: p = M$

Fig. 3: Self and cross difference sets for CACIS configuration with $M=4$, $N=3$.

$$\mathcal{L}^+_{S\breve{M}} = \{l_s \mid l_s = \breve{M}n\}$$
$$\mathcal{L}^+_{SN} = \{l_s \mid l_s = Nm\} \qquad (2)$$

where $0 \leq n \leq N$-$1$ and $0 \leq m \leq M$-$1$. The corresponding mirrored position set is denoted by $\mathcal{L}^-_{S\breve{M}}$ and $\mathcal{L}^-_{SN}$. The union of the self-difference sets is given by $\mathcal{L}_S = \mathcal{L}^+_S \cup \mathcal{L}^-_S = \mathcal{L}^+_{S\breve{M}} \cup \mathcal{L}^+_{SN} \cup \mathcal{L}^-_{S\breve{M}} \cup \mathcal{L}^-_{SN}$. The analysis of the self-difference sets for the prototype co-prime array presented in [5, 20] is applicable even for the CACIS configuration with inter-element spacing of $\breve{M}d$, since the range of $m$ is the same as that of the prototype co-prime array (Fig. 3).

Cross differences are the difference values or lags generated between the first and second sub-array. The cross-difference set $\mathcal{L}^+_C$ is given by:

$$\mathcal{L}^+_C = \{l_c \mid l_c = \breve{M}n - Nm\} \qquad (3)$$

where $0 \leq n \leq N$-$1$ and $0 \leq m \leq M$-$1$. Set $\mathcal{L}^+_C$ and its mirrored position set $\mathcal{L}^-_C$ have $MN$ unique differences but the number of unique differences in the union set $\mathcal{L}_C = \mathcal{L}^+_C \cup \mathcal{L}^-_C$ differs from that of the prototype co-prime array (remember prototype is a special case of the CACIS).

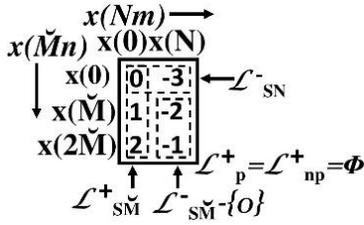

Fig. 4: Cross difference set for *M=2, N=3, p=2, M̆=1*.

The analysis of the cross difference set for CACIS scheme is not straightforward and hence, to get a better insight into the cross difference set, some additional sets are defined: $\mathcal{L}^+_p$, $\mathcal{L}^-_p$, $\mathcal{L}^+_{np}$, $\mathcal{L}^-_{np}$, $\mathcal{L}_p$ and $\mathcal{L}_{np}$. The set containing mirrored pairs is defined as:

$$\mathcal{L}^+_p = \{l_c | l_c = \breve{M}n - Nm, n\epsilon[1, N-1], m \epsilon[1, \breve{M} - 1]\} \quad (4)$$

$\mathcal{L}^-_p$ is the set containing values that are negative of the values in $\mathcal{L}^+_p$. $\mathcal{L}^+_p$ satisfies the following condition which also holds true for $\mathcal{L}^-_p$:

$$\{-l_c \in \mathcal{L}^+_p | l_c \in \mathcal{L}^+_p\}$$

The set containing no mirrored pairs is defined as:

$$\mathcal{L}^+_{np} = \{l_c | l_c = \breve{M}n - Nm, n \epsilon [1, N-1], m \epsilon [\breve{M} + 1, M - 1]\} \quad (5)$$

$\mathcal{L}^-_{np}$ is the mirrored position set of the differences contained in $\mathcal{L}^+_{np}$. Let us also define two new union sets $\mathcal{L}_p$ and $\mathcal{L}_{np}$ as:

$$\mathcal{L}_p = \mathcal{L}^+_p \cup \mathcal{L}^-_p$$

$$\mathcal{L}_{np} = \mathcal{L}^+_{np} \cup \mathcal{L}^-_{np} \quad (6)$$

For the case of the prototype co-prime array (*p=1*), the sets $\mathcal{L}^+_{np}$, $\mathcal{L}^-_{np}$, and $\mathcal{L}_{np}$ are empty sets. Since $\breve{M} = M$, the range of *m* defined in (5) does not hold. Therefore, these sets do not exist. Here, it may be noted that $\mathcal{L}^-_{S\breve{M}} - \{0\} \nsubseteq \mathcal{L}^+_C, \mathcal{L}^+_{S\breve{M}} - \{0\} \nsubseteq \mathcal{L}^-_C, \mathcal{L}^+_p = \mathcal{L}^+_C - \mathcal{L}_S$ and $\mathcal{L}^-_p = \mathcal{L}^-_C - \mathcal{L}_S$. The cross difference set in this case corresponds to (Fig. 3 in [5]).

When *p* lies in the range $2 \leq p \leq M-1$, none of the sets are empty and this is depicted in Fig. 3(c)-3(d) for the case when *p=2*.

On the other hand, for the nested array (*p=M*), the sets $\mathcal{L}^+_p, \mathcal{L}^-_p$, and $\mathcal{L}_p$ are empty sets. Since $\breve{M}=1$, the range of *m* in (4) does not hold. The cross difference sets for the nested array is shown in Fig. 3(e)-3(f).

Proposition I:

1. For $l_c$ belonging to set $\mathcal{L}^+_p$

$$\{l_c | l_c < 0, l_c \in \mathcal{L}^+_p\} \subseteq \{-l_c | l_c > 0, l_c \in \mathcal{L}^+_C\}$$

$$\{l_c | l_c < 0, l_c \in \mathcal{L}^+_p\} \subseteq \{-l_c | l_c > 0, l_c \in \mathcal{L}^+_C - \mathcal{L}_S - \mathcal{L}_{np}\}$$

2. For $l_c$ belonging to set $\mathcal{L}^-_p$

$$\{l_c | l_c < 0, l_c \in \mathcal{L}^-_p\} \subseteq \{-l_c | l_c > 0, l_c \in \mathcal{L}^-_C\}$$

$$\{l_c | l_c < 0, l_c \in \mathcal{L}^-_p\} \subseteq \{-l_c | l_c > 0, l_c \in \mathcal{L}^-_C - \mathcal{L}_S - \mathcal{L}_{np}\}$$

3. For $l_c$ belonging to set $\mathcal{L}^+_{np}$
$$\{l_c < 0, \quad \forall \, l_c \in \mathcal{L}^+_{np}\}$$

4. For $l_c$ belonging to set $\mathcal{L}^+_{np}$

$$\{l_c | l_c < 0, l_c \in \mathcal{L}^+_{np}\} \nsubseteq \{-l_c | l_c > 0, l_c \in \mathcal{L}^+_C\}$$

5. For $l_c$ belonging to set $\mathcal{L}^-_{np}$
$$\{l_c > 0, \quad \forall \, l_c \in \mathcal{L}^-_{np}\}$$

6. For $l_c$ belonging to set $\mathcal{L}^-_{np}$

$$\{l_c | l_c > 0, l_c \in \mathcal{L}^-_{np}\} \nsubseteq \{-l_c | l_c < 0, l_c \in \mathcal{L}^-_C\}$$

The claims in Proposition I are proved in Appendix A.

## 4. UNIQUENESS, CONTINUITY AND NUMBER OF CONTRIBUTORS

Let us now estimate the degrees of freedom (dof), (which is also referred to as the number of unique or distinct values) for each of the sets defined in the previous section.

Proposition II:

1. Sets $\mathcal{L}^+_p$ and $\mathcal{L}^-_p$ contain ($\breve{M}$-1)(N-1) unique differences.
2. Sets $\mathcal{L}^+_{np}$ and $\mathcal{L}^-_{np}$ contain (N-1)(M-$\breve{M}$-1) unique differences.
3. Set $\mathcal{L}_p$ contains ($\breve{M}$-1)(N-1) unique differences.
4. Set $\mathcal{L}_{np}$ contains 2(N-1)(M-$\breve{M}$-1) unique differences.
5. Sets $\mathcal{L}^+_C$ and $\mathcal{L}^-_C$ are given by:

$$\mathcal{L}^+_C = \mathcal{L}^+_{S\breve{M}} \cup \mathcal{L}^-_{SN} \cup \mathcal{L}^+_p \cup \mathcal{L}^+_{np} \cup [\mathcal{L}^-_{S\breve{M}} - \{0\}]$$

$$\mathcal{L}^-_C = \mathcal{L}^-_{S\breve{M}} \cup \mathcal{L}^+_{SN} \cup \mathcal{L}^-_p \cup \mathcal{L}^-_{np} \cup [\mathcal{L}^+_{S\breve{M}} - \{0\}]$$

6. Set $\mathcal{L}_C = \mathcal{L}^+_C \cup \mathcal{L}^-_C$ has $N(2M-1) - \breve{M}(N-1)$ unique differences for $2 \leq p \leq M-1$. The general expression for the number of unique differences is given by:

$$2(M + N - 1) - 1 + \left\lceil \frac{M-p}{M} \right\rceil (\breve{M} - 1)(N - 1) + \left\lceil \frac{p-1}{M} \right\rceil 2(N - 1)(M - \breve{M} - 1) \quad (7)$$

The claims in Proposition II are proved in Appendix B. It is evident from Propositions II-5 and II-6 that the self-differences are a subset of the cross-differences. Hence, the union of all the sets, $\mathcal{L}$, is equal to the cross difference set i.e. $\mathcal{L} = \mathcal{L}^+_C \cup \mathcal{L}^-_C$. A summary of the degrees of freedom or the number of unique differences in each set is given in Table 1.

For the CACIS scheme, Proposition III gives the range of integers in the difference set and the continuous range (without holes or missing values).

Proposition III:

1. $\mathcal{L}^+_C$ has MN distinct integers in the range $-N(M-1) \leq l_c \leq \breve{M}(N-1)$
2. $\mathcal{L}^-_C$ has MN distinct integers in the range $-\breve{M}(N-1) \leq l_c \leq N(M-1)$
3. $\mathcal{L}^+_C$ has consecutive integers in the range $-MN + \breve{M}(N-1) + 1 \leq l_c \leq N-1$
4. $\mathcal{L}^-_C$ has consecutive integers in the range $-(N-1) \leq l_c \leq MN - \breve{M}(N-1) - 1$
5. $\mathcal{L}_C$ has consecutive integers in the range $-MN + \breve{M}(N-1) + 1 \leq l_c \leq MN - \breve{M}(N-1) - 1$ which implies that this set has its first hole at $|l_c| = MN - \breve{M}(N-1)$ when $2 \leq p \leq M$.

Proposition III-5 is not valid for $p=1$ since this situation implies $M=\breve{M}$ which yields $MN - \breve{M}(N-1) = M = \breve{M}$ which cannot be a hole in the set $\mathcal{L}$ since it is a self-difference. The range in this case was given by Proposition II-(7) in [5]. A discussion on Proposition III is provided in Appendix C.

The number of sample pairs whose cross-difference maps to the same difference value $l$ is a critical factor in the estimation process. It is also referred to as the weight function. The larger the number of sample pairs, the more the accurate the estimate of autocorrelation at that difference value would be. Let the number of samples contributing to the estimation of the autocorrelation at each difference value $l \in \mathcal{L}$ be denoted by $z(l)$.

Proposition IV:

1. For $l \in \{\mathcal{L}^+_{S\breve{M}} \cup \mathcal{L}^-_{S\breve{M}}\}$ excluding the difference value of zero:

$$z(l) = (N - i) + \left\lceil \frac{p-1}{M} \right\rceil,$$

for, $\{1 \leq i \leq N-1, l = \pm \breve{M}i\}$.

2. For $l \in \{\mathcal{L}^+_{SN} \cup \mathcal{L}^-_{SN}\}$ excluding the difference value of zero:

$$z(l) = (M - i)$$
for, $\{1 \leq i \leq M-1, l = \pm Ni\}$.

3. For the difference value of zero:
$$z(l) = M + N - 1$$
4. For $l \in \mathcal{L}^+_p$ and $\mathcal{L}^-_p$: $z(l) = 2$.
5. For $l \in \mathcal{L}_p$: $z(l) = 2$.
6. For $l \in \mathcal{L}^+_{np}, \mathcal{L}^-_{np}$ and $\mathcal{L}_{np}$: $z(l) = 1$.

Propositions IV-1, IV-2 and IV-3 are similar to the prototype co-prime array described in [5, 20], except that Proposition IV-1 has an additional term $\left\lceil \frac{p-1}{M} \right\rceil$ which arises due to the fact that for $2 \leq p \leq M$ the cross difference sets $\mathcal{L}^+_C$ and $\mathcal{L}^-_C$ have elements from $\mathcal{L}^-_{S\breve{M}} - \{0\}$ and $\mathcal{L}^+_{S\breve{M}} - \{0\}$ respectively, for $m=\breve{M}$ and $n \in [1, N-1]$.

The number of unique differences (Proposition II-1) in the set $\mathcal{L}^+_p$ (and $\mathcal{L}^-_p$) is equal to the maximum number of elements in it. Therefore, each difference value occurs once, i.e. $z(l)$ should be equal to one. But under the assumption that the signal is wide sense stationary, knowledge at lag $l$ and $-l$ is equivalent. Proposition I-1 (and Fig. 3(c)) shows that $(l, -l)$ exist as a pair in $\mathcal{L}^+_p$. Thus, justifying Proposition IV-4. For Proposition IV-5, the sum of the cardinality of the paired sets, $|\mathcal{L}^+_p|+|\mathcal{L}^-_p|=2(\breve{M}-1)(N-1)$, but the total number of unique differences in $\mathcal{L}_p = \mathcal{L}^+_p \cup \mathcal{L}^-_p = (\breve{M}-1)(N-1)$. Therefore, there are two pairs, $(m, n)$, that map to each element in the difference set $\mathcal{L}_p$.

Proposition IV-6 follows from Proposition II-2 since the maximum possible combinations for the pair $(m, n)$ in the set $\mathcal{L}^+_{np}$ and $\mathcal{L}^-_{np}$ is $(N-1)(M-\breve{M}-1)$ which also corresponds to the number of unique differences. In addition, these sets do not contain mirrored pairs. The proof for the set $\mathcal{L}_{np}$ follows from Proposition II-4 since the maximum possible combinations for the pair $(m, n)$ in the set $\mathcal{L}_{np}$ is $2(N-1)(M-\breve{M}-1)$ which also corresponds to the number of unique differences. Therefore, only one contributor is available for estimation.

Table 1: Summary of unique differences or *dof* for each difference set.

| Set | $\mathcal{L}^+_{S\breve{M}}$ and $\mathcal{L}^-_{S\breve{M}}$ | $\mathcal{L}^+_{SN}$ and $\mathcal{L}^-_{SN}$ | $\mathcal{L}^+_S$ and $\mathcal{L}^-_S$ | $\mathcal{L}_S$ | $\mathcal{L}^+_C$ and $\mathcal{L}^-_C$ | $\mathcal{L}_C$ and $\mathcal{L}$ |
|---|---|---|---|---|---|---|
| *dof* | N | M | M+N-1 | 2(M+N-1)-1 | MN | $N(2M-1) - \breve{M}(N-1)$ |

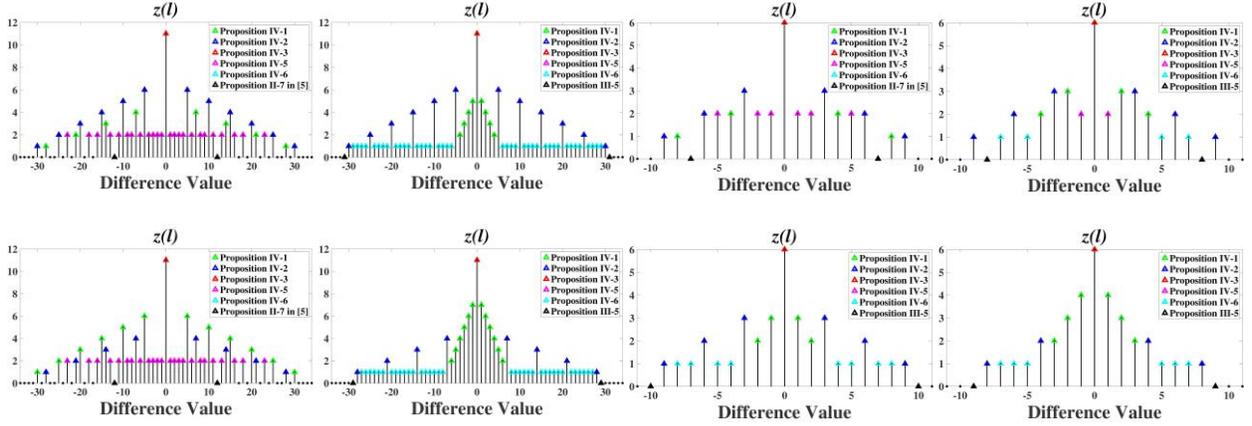

Fig. 5: Weight function: top-left (*M=7*, *N=5*, *p=1*), top-right (*M=7*, *N=5*, *p=7*), bottom-left (*M=5*, *N=7*, *p=1*), bottom-right (*M=5*, *N=7*, *p=5*).

Fig. 6: Weight function: top-left (*M=4*, *N=3*, *p=1*), top-right (*M=4*, *N=3*, *p=2*), bottom-left (*M=4*, *N=3*, *p=4*), bottom-right (*M=3*, *N=4*, *p=3*).

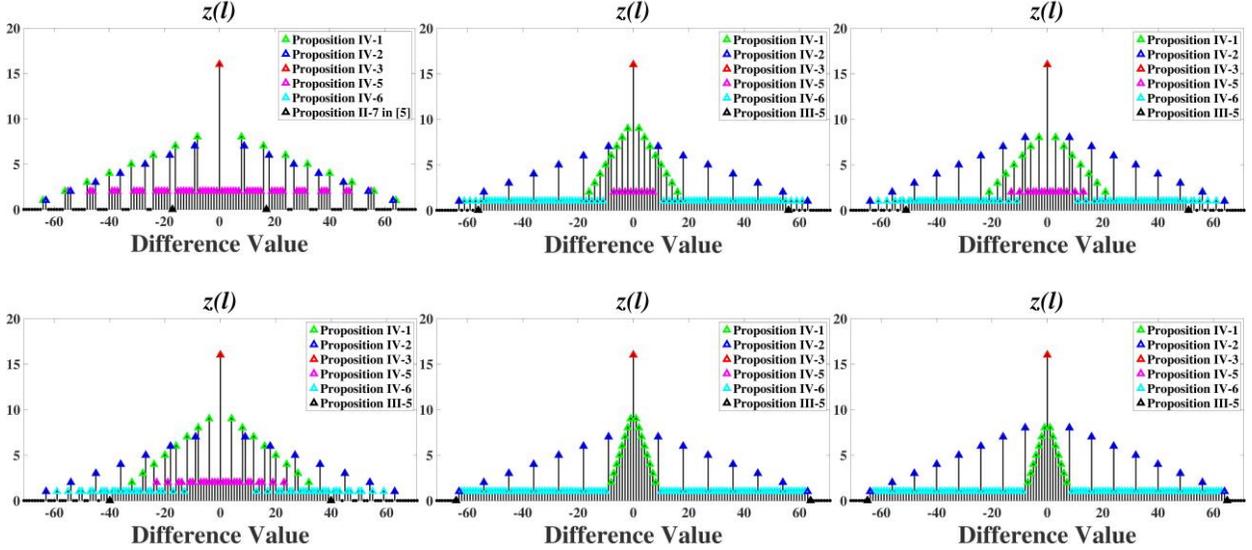

Fig. 7: Weight function: top-left (*M=8*, *N=9*, *p=1*), bottom-left (*M=8*, *N=9*, *p=2*), top-middle (*M=8*, *N=9*, *p=4*), bottom-middle (*M=8*, *N=9*, *p=8*), top-right (*M=9*, *N=8*, *p=3*), bottom-right (*M=9*, *N=8*, *p=9*).

## 5. DISCUSSION

The fundamentals of the difference set for the CACIS configuration has been developed in the preceding sections. The focus here is to provide more insights into the theory by using several numerical examples. We will primarily focus on the weight function for different values of *M*, *N*, and *p*, since it contains most of the critical information that governs the usefulness of the CACIS scheme.

Let us consider a CACIS configuration with parameters $M=7$, $N=5$ and $\breve{M} = \frac{M}{p}$, where $\breve{M}$ and $N$ are co-prime. So, what are the possible values of $p$ that can be used in this case? We know that $1 \leq p \leq M$, therefore, the choice is limited to $\{1, 2, 3, 4, 5, 6, 7\}$. It may be noted that only $p=1$ (prototype co-prime array) and $p=7$ (nested array) are valid options since for other values $\breve{M}$ is not an integer. These two cases are shown in Fig.5 where the extreme non-zero values on the x-axis represent the range of difference values in the union set. The number of non-zero weights represent the number of unique difference values or the degrees of freedom in the union set given by (7) (Proposition II-6). As shown in Fig. 5, for the case when $M=7$ and $N=5$ the degrees of freedom are 45 and 61 with $p=1$ and $p=7$ respectively, and is in accordance with the derived formula. The weight at each difference value corresponds to Proposition IV and represents the number of contributors for autocorrelation estimation. These parameters influence the accuracy, convergence and latency of the scheme. The first zero on either side is located at 12 and 31 for $p=1$ (Proposition II-7 in [5]) and for $p=7$ (Proposition III-5) respectively. It provides the continuous range of difference values for autocorrelation estimation. Fig. 5 also contains the weight function for the case when $M$ and $N$ are reversed i.e. $M=5$ and $N=7$. This scenario also has two possibilities with $p=1$ and $p=5$. It may be noted that interchanging the value of $M$ and $N$ with $p=1$ does not alter the weight function. This is expected since the prototype co-prime array has the same properties on interchanging the values of $M$ and $N$. However, the weight function changes for values of $p \neq 1$ as in Fig.5 (top-right) and (bottom-right). It is important to note that the choice of $M$ decides the possible values of $p$ and selecting a prime number limits the value of $p$ to 1 and M.

Therefore, let us consider another example with $M=4$ and $N=3$, which has been used to describe the difference sets in the preceding sections. The possible values of $p$ are $\{1, 2, 4\}$, and its weight function as derived in Proposition IV is shown in Fig. 6. It also contains the weight function for the case when $M$ and $N$ are interchanged, i.e. $M=3$ and $N=4$. Here the compression factor can take a value of $p=1$ or $p=3$. Since $p=1$ has the same weight function for (M=4, N=3) and (M=3, N=4), it has been shown only once. The first hole is located at 7, 8, 10 and 9 for (M, N, p) equal to (4, 3, 1), (4, 3, 2), (4, 3, 4), and (3, 4, 3) respectively, and is in line with Proposition III-5. The degrees of freedom for the above four scenarios are $\{17, 17, 19, 17\}$ respectively and is in accordance with Proposition II-6 (equation (7)).

Next, we consider an example with $M=8$ and $N=9$, with $p = \{1, 2, 4, 8\}$. The location of the first hole is given by $\{17, 40, 56, 64\}$ and the number of unique difference values that can be realized is given by $\{87, 103, 119, 127\}$. For the case when $M=9$ and $N=8$, we have $p = \{1, 3, 9\}$ with the location of the first zero given by $\{17, 51, 65\}$ and the number of unique differences given by $\{87, 115, 129\}$. The theoretical values match the plots depicted in Fig. 7. All the examples considered here validates the theoretical expressions developed in this paper.

## 6. BIAS OF CORRELOGRAM

The correlogram technique for spectral estimation has an inherent bias which is given by the Fourier transform of the weight function. This is well studied for the Nyquist case in [22, 23] and was derived for the prototype co-prime array in [20]. It also describes the correlogram method when an unbiased and a biased autocorrelation estimate is employed. The importance of this bias window expression is that its convolution with the true spectrum gives an approximation of the co-prime spectrum. This can therefore be the basis for undoing the distortion.

Here, the mathematical expressions (closed-form) for the weight function and bias window of the correlogram estimate for the CACIS scheme is given by (8) and (9) respectively. These expressions are developed for the entire difference set including holes and for the case when a biased autocorrelation estimate is used for spectral estimation. Note that weight function (8) is obtained from Proposition IV.
The bias window (9) is the Fourier transform of (8). We verify the correctness of the derived bias expression (9) in Fig. 8-10 by comparing it with the simulated bias window obtained as the FFT of the weight function, for examples considered in Fig. 5-7 respectively.

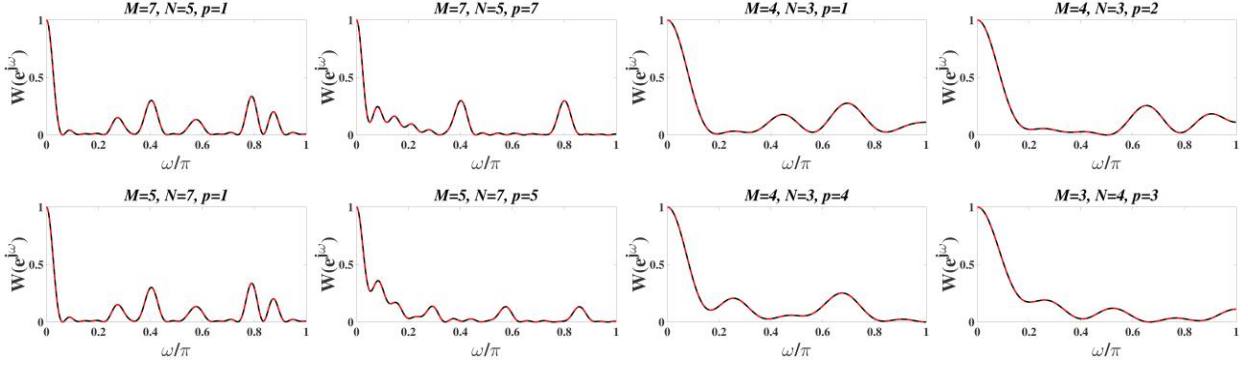

Fig. 8: Bias window for $(M, N) = (7, 5)$ and $(5, 7)$.

Fig. 9: Bias window for $(M, N) = (4, 3)$ and $(3, 4)$.

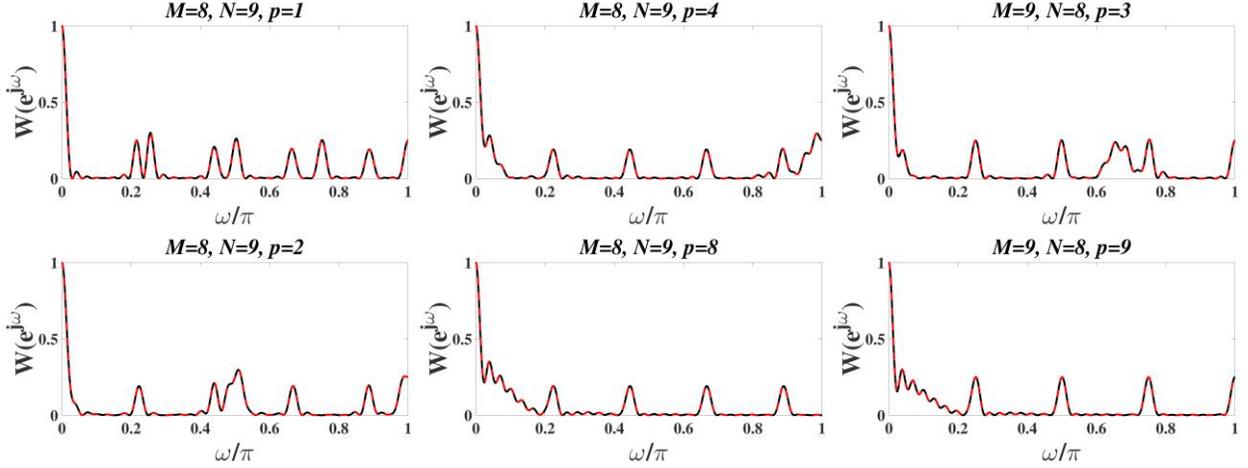

Fig. 10: Bias window for $(M, N) = (8, 9)$ and $(9, 8)$.

$$z(l) = \sum_{n=-N+1}^{N-1} \left(N - |n| + \left\lceil \frac{p-1}{M} \right\rceil\right) \delta(l - \widetilde{M}n) - \left(1 + \left\lceil \frac{p-1}{M} \right\rceil\right) \delta(l) + \sum_{m=-M+1}^{M-1} (M - |m|) \delta(l - Nm)$$
$$+ \sum_{n=1}^{N-1} \sum_{m=1}^{\widetilde{M}-1} 2\delta\left(l - (\widetilde{M}n - Nm)\right) + \sum_{n=1}^{N-1} \sum_{m=\widetilde{M}+1}^{M-1} \delta(|l| - (Nm - \widetilde{M}n))$$
(8)

$$W(e^{j\omega}) = \frac{1}{s}\left\{\left|\frac{\sin\frac{\omega\widetilde{M}N}{2}}{\sin\frac{\omega\widetilde{M}}{2}}\right|^2 + \left|\frac{\sin\frac{\omega MN}{2}}{\sin\frac{\omega N}{2}}\right|^2 + 2\frac{\sin\frac{\omega\widetilde{M}(N-1)}{2}\sin\frac{\omega N(\widetilde{M}-1)}{2}}{\sin\frac{\omega\widetilde{M}}{2}\sin\frac{\omega N}{2}}\right.$$
$$\left. + 2\cos\left(\frac{\omega MN}{2}\right)\frac{\sin\frac{\omega\widetilde{M}(N-1)}{2}\sin\frac{\omega N(M-\widetilde{M}-1)}{2}}{\sin\frac{\omega\widetilde{M}}{2}\sin\frac{\omega N}{2}} + \frac{\sin\frac{\omega\widetilde{M}(2N-1)}{2}}{\sin\frac{\omega\widetilde{M}}{2}} - 2\right\}$$
(9)

In Fig. 8, the bias for $(M, N, p) = (7, 5, 1)$ and $(5, 7, 1)$ are identical since their weight functions match. It is observed that the bias for $(7, 5, 7)$ and $(5, 7, 5)$ have ripples in the vicinity of the main lobe and the local minima of the ripples are non-zero. This can cause spectral leakage in the vicinity of the true spectral peak. In Fig. 9, for $M=4$, $N=3$ and vice versa, we observe fewer ripples in the vicinity of the main lobe as well as fewer side-lobes since the parameters $(M, N, p)$ are smaller when compared to the previous example, however, the main lobe width is larger.

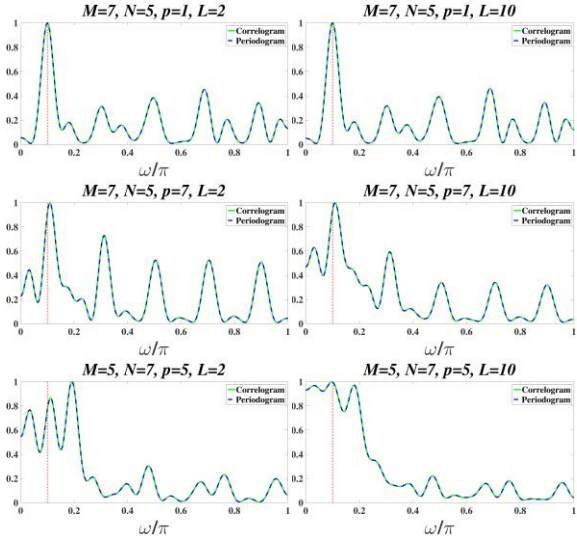

Fig. 11: Single spectral peak estimation for $(M, N) =$ (7, 5) and (5, 7).

In Fig. 10, the ripples at the base and the number of side-lobes are more as compared to the previous two examples since the parameters $(M, N, p)$ is large. This case with $M=8$, $N=9$ and vice versa, also has a narrow main-lobe. From all the examples considered, it may be noted that the prototype co-prime array with $p=1$ is the only case in which the main-lobe of the bias drops to near zero without any ripple at the base, and as the value of $p$ increases the ripples tend to increase.

## 7. SIMULATIONS

Let us now analyze the CACIS scheme for temporal spectral estimation of a signal in the absence of noise. We begin with a signal model which contains a single spectral peak at $0.1\pi$. This analysis is necessary to judge the usefulness of the CACIS configuration for different parameters $(M, N, p)$ since the bias window for larger values of $p$ has ripples at the base of the main lobe. The correlogram-based spectral estimate is shown in Fig. 11-13 for the examples considered in Fig. 8-10. Note that $L$ represents the number of snapshots. $L=2$ and $L=10$ (low latency) have been considered here. $(M, N, p)=(7, 5, 1)$ gives a clean estimate of the peak at $0.1\,\Pi$ in Fig. 11, while the estimate has spurious peaks in the vicinity of $0.1\pi$ for parameters (7, 5, 7) and (5, 7, 5) with (5, 7, 5) totally failing since the peak at $0.1\pi$ is approximately equal to (if not lower than) the peaks in the vicinity.

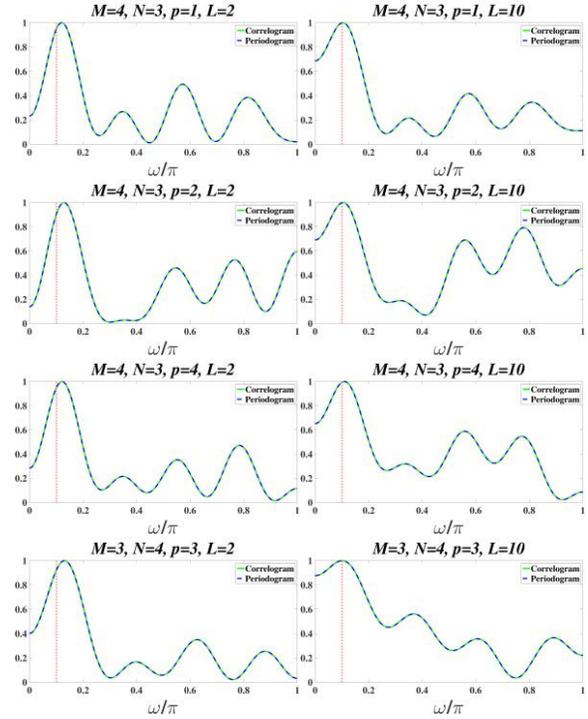

Fig. 12: Single spectral peak estimation for $(M, N) =$ (4, 3) and (3, 4).

Single peak spectral estimation when $(M, N)=(4, 3)$ and (3, 4) is shown in Fig. 12. Few large side-lobes appear away from the true peak location but are lower in amplitude than the true peak. Fig. 13 shows that (8, 9, 1) and (8, 9, 2) provides a good estimate of the peak. It is also noted that despite some distortion, (8, 9, 4) and (8, 9, 8) is able to estimate the true peak; while (9, 8, 3) and (9, 8, 9) fails to provide a reliable estimate of the spectral peak.

For most of the examples considered, it is observed that the distortion in the vicinity of the spectral peak is least for $p=1$. This can be attributed to low ripples at the base of the bias window (refer Fig. 8-10). Also note that the periodogram and correlogram estimates match well. In a statistical sense, the correlogram spectral estimate is the convolution of the true spectrum with the derived bias window. Therefore, we wish to have a bias window with narrow main lobe width, low ripples at the vicinity of the main lobe, and low side-lobes.

Next, we analyze the ability of the CACIS scheme for estimating multiple peaks located at $0.1\pi$, $0.3\pi$ and $0.6\pi$ in a noise-free environment.

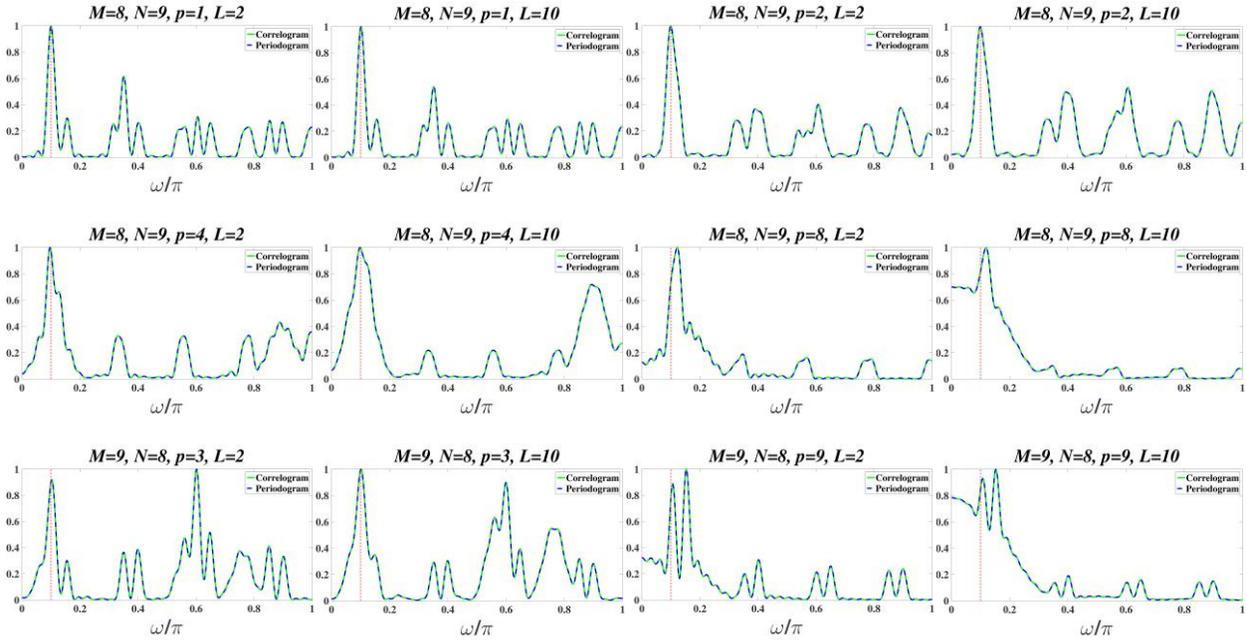

Fig. 13: Single spectral peak estimation for $(M, N) = (8, 9)$ and $(9, 8)$.

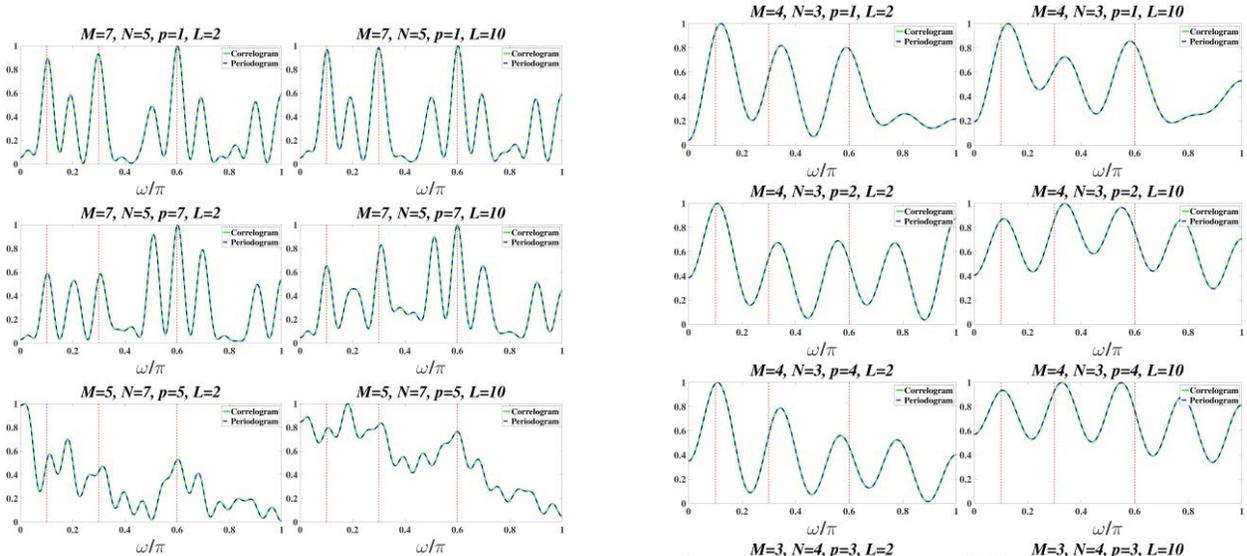

Fig. 14: Multiple spectral peak estimation for $(M, N) = (7, 5)$ and $(5, 7)$.

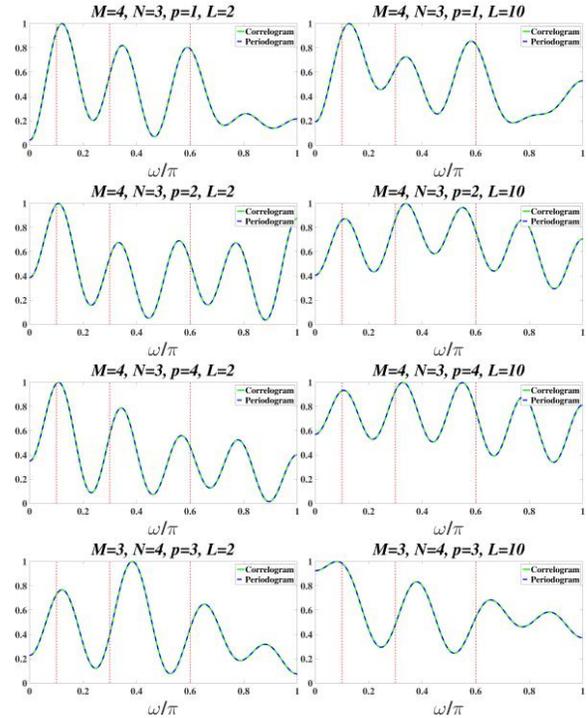

Fig. 15: Multiple spectral peak estimation for $(M, N) = (4, 3)$ and $(3, 4)$.

The signal model considered is similar to that in [20]. As shown in Fig. 14 the prototype co-prime array with parameters (7, 5, 1) accurately estimates the three peaks while (7, 5, 7) and (5, 7, 5) fail. In Fig. 15, we observe that (4, 3, 1) (prototype co-prime array) and (3, 4, 3) seem to detect the three spectral peaks, but the frequency locations of the peaks are not accurate. However, a fourth peak is observed for (3, 4, 3) and has an amplitude comparable with the third peak which can lead to an error in peak detection. In Fig.16, the correlogram and periodogram method estimates one or at most two peaks with many large spurious peaks causing the estimation process to fail. It is important to note that the ability of the CACIS scheme to estimate multiple spectral peaks depends on the shape of the bias window. In general, CACIS could be designed with appropriate values of (M, N, p) to minimize the effect of bias distortion. In addition, variants of the standard correlogram method can be investigated in the future.

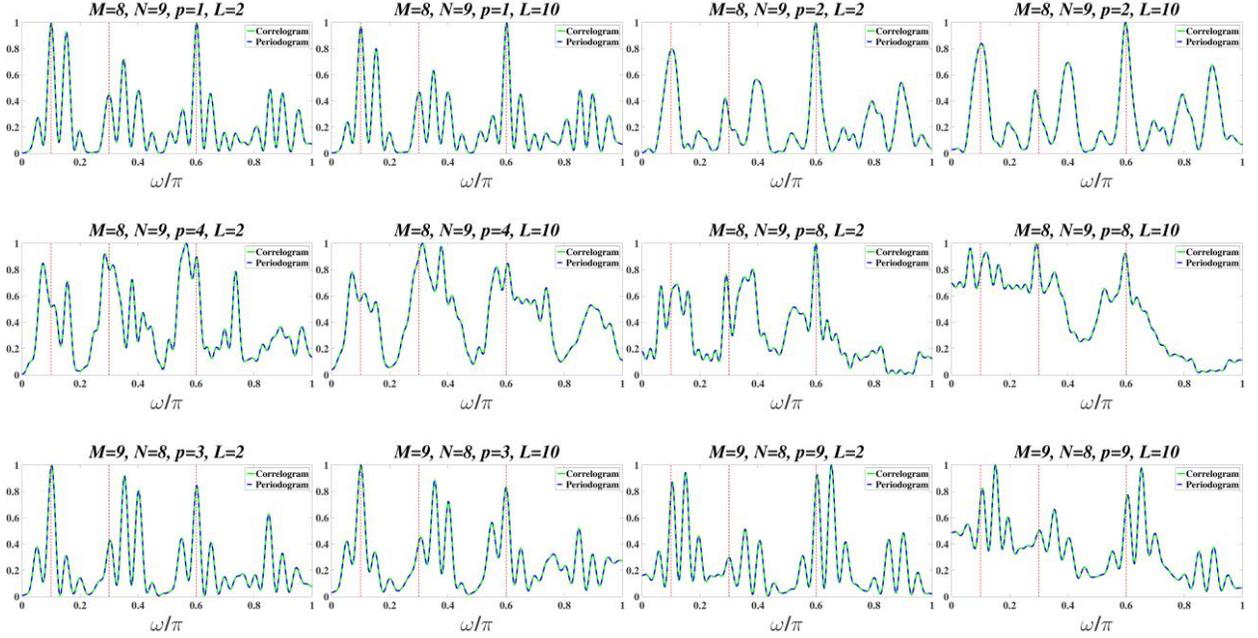

Fig. 16: Multiple spectral peak estimation for $(M, N) = (8, 9)$ and $(9, 8)$.

## 8. CONCLUSION

This paper studies the correlogram spectral estimator. It makes no assumptions on the signal model and achieves low latency estimation. However, ripples are observed in the bias window and spectral estimate. This may not be a good sign, but researchers can work on mitigating it. Therefore, correlogram may not appear useful for DoA estimation which has been studied by researchers using subspace-based algorithms with better accuracy. However, the low latency achieved in this paper may prompt researchers to study the subspace-based algorithms to achieve lower latency.

The fundamentals developed can form the basis for the variance and Cramér-Rao bound analysis. Other qualitative measures like mean-square error, peak location, etc. may be considered. Optimization of the parameters (M, N, p) to achieve objectives like minimum bias distortion, minimum variance, large continuous range, etc. can be investigated based on the application. There are several other co-prime based structures whose difference set for low latency estimation and closed-form expressions can be derived along similar lines. Some of the structures are n-tuple or multi-level prime arrays [24, 25], thinned co-prime arrays [26], multiple period co-prime arrays [27], etc. In addition, the CACIS scheme may be analyzed under perturbed conditions as considered in [28-30]. The effect of window functions on the estimate may be considered for this scheme as noted in [20, 31].

## APPENDIX A

### PROOF OF PROPOSITION I

1. Let $l_{c1} = \breve{M} n_1 - N m_1$ be an element of set $\mathcal{L}^+_p$ where $n_1 \in [1, N-1]$, $m_1 \in [1, \breve{M}-1]$, and $l_{c2} = \breve{M} n_2 - N m_2$ be an element of set $\mathcal{L}^+_c$. Let us assume there exists $(n_2, m_2)$ such that $l_{c2} = -l_{c1}$. This implies $\breve{M} n_2 - N m_2 = N m_1 - \breve{M} n_1$ and hence, $\frac{\breve{M}}{N} = \frac{m_1 + m_2}{n_1 + n_2}$. Since $\breve{M}$ and N are co-prime, the above equation will hold if and only if $m_1 + m_2 = \breve{M}$ and $n_1 + n_2 = N$. Thus, $m_2 = \breve{M} - m_1$ and $n_2 = N - n_1$.

  Substituting the range of $m_1$ and $n_1$ in the equation for $m_2$ and $n_2$, we get $m_2 \in [1, \breve{M} - 1]$

and $n_2 \in [1, N-1]$. This implies that $l_{c2} = -l_{c1}$ does hold for all values of $l_{c1} \in \mathcal{L}^+_p$. It is easy to see that the range of $(m, n)$ for set $\mathcal{L}^+_p$ is also contained in $\mathcal{L}^+_C$ and $\mathcal{L}^+_C - \mathcal{L}_S - \mathcal{L}_{np}$, thus proving the claims in Proposition I-1. In simple terms, this implies that $\mathcal{L}^+_p$ has mirrored pairs, e.g. [1, -1] (refer Fig. 3(c)).

2. Proposition I-2 has a similar proof as Proposition I-1.

3. Let $l_c \in \mathcal{L}^+_{np}$, with $m \in [\breve{M}+1, M-1]$ and $n \in [1, N-1]$. Substitute $(m, n)$ in $l_c = \breve{M}n - Nm$ to find the minimum and maximum value of $l_c$. This leads to the range $-MN+N+\breve{M} \leq l_c \leq -(\breve{M}+N)$. Since the upper limit is guaranteed to be negative the lower limit would obviously be negative. Specifically, for the lower limit to be negative we need to satisfy $(\breve{M}+N) < MN$. The set $\mathcal{L}^+_{np}$ is not defined for the prototype co-prime array, hence, $\breve{M} < M$. This proves the claim made.

4. Let $l_{c1} = \breve{M}n_1 - Nm_1$ be an element of set $\mathcal{L}^+_{np}$ with $m_1 \in [\breve{M}+1, M-1]$ and $n_1 \in [1, N-1]$. Let $l_{c2} = \breve{M}n_2 - Nm_2$ be an element of set $\mathcal{L}^+_C$ with $m_2 \in [0, M-1]$ and $n_2 \in [0, N-1]$.
Let us assume that there exists some $l_{c2} = -l_{c1}$. This implies $\breve{M}n_2 - Nm_2 = Nm_1 - \breve{M}n_1$ and hence, $\frac{\breve{M}}{N} = \frac{m_1+m_2}{n_1+n_2}$, which will hold if and only if $m_1 + m_2 = \breve{M}$ and $n_1 + n_2 = N$. Thus, $m_2 = \breve{M} - m_1$ and $n_2 = N - n_1$. Substituting the ranges of $m_1$ and $n_1$ in the equation for $m_2$ and $n_2$ we get, $(1-p)\breve{M}+1 \leq m_2 \leq -1$ and $1 \leq n_2 \leq N-1$. Though the range for $n_2$ is satisfied, that of $m_2$ is not. Since the upper limit of $m_2$ is negative, it is obvious that the lower limit will be negative. Specifically, since $p \in [2, M]$ it can be shown that $(1-p)\breve{M}+1$ can take a maximum value of $-\breve{M}+1$ for $p=2$ which is guaranteed to be negative. This guarantee appears to fail in the case when $p=2$ and $\breve{M}=1$ which implies $M=p\breve{M}=2$. But this case will never arise since $\mathcal{L}^+_{np}$ is not defined for this scenario and is an empty set as shown in Fig. 4. This implies that the difference values in the set $\mathcal{L}^+_{np}$ do not contain a mirrored counterpart within $\mathcal{L}^+_C$. But it has a mirrored counterpart in set $\mathcal{L}^-_C$ since $\mathcal{L}^-_{np} \subseteq \mathcal{L}^-_C$. As an example, Fig. 3(c) has $l_c = [-7, -5] \in \mathcal{L}^+_{np}$ with no mirrored pair, i.e. $[+7, +5]$ in $\mathcal{L}^+_C$.

5. Proposition I-5 has a similar proof as Proposition I-3.

6. Proposition I-6 can be proven similar to Proposition I-4 by using $l_{c1} = Nm_1 - \breve{M}n_1$ and $l_{c2} = Nm_2 - \breve{M}n_2$ as elements of the sets $\mathcal{L}^-_{np}$ and $\mathcal{L}^-_C$, respectively.

## APPENDIX B

### PROOF OF PROPOSITION II

1. Let $l_{c1} = \breve{M}n_1 - Nm_1$ and $l_{c2} = \breve{M}n_2 - Nm_2$ be two elements of set $\mathcal{L}^+_p$.
Let us assume that there exists some $(n_2, m_2)$ such that $l_{c2} = l_{c1}$. This implies that $\breve{M}n_2 - Nm_2 = \breve{M}n_1 - Nm_1$ and hence, $\frac{\breve{M}}{N} = \frac{m_1-m_2}{n_1-n_2}$. Since $|m1 - m2| < \breve{M}$, $|n1 - n2| < N$ and $(\breve{M}, N)$ are co-prime, the above equation cannot be satisfied. This implies that every element in the set $\mathcal{L}^+_p$ is unique and the total number of unique differences equal the number of possible combinations of $(m, n)$ i.e. $(\breve{M}-1)(N-1)$. Since the elements in $\mathcal{L}^-_p$ are mirrored version of the elements in $\mathcal{L}^+_p$, the same holds true for $\mathcal{L}^-_p$.

2. Let $l_{c1} = \breve{M}n_1 - Nm_1$ and $l_{c2} = \breve{M}n_2 - Nm_2$ be two elements of set $\mathcal{L}^+_{np}$. Let us assume that there exists some $(n_2, m_2)$ such that $l_{c2} = l_{c1}$. This implies, $\frac{\breve{M}}{N} = \frac{m_1-m_2}{n_1-n_2}$. Since $|n1 - n2| < N$ and $(\breve{M}, N)$ are co-prime, the above equation cannot be satisfied. This implies that every element in the set $\mathcal{L}^+_{np}$ is unique and the total number of unique differences equal the number of possible combinations of $(m, n)$ i.e. $(N-1)(M-\breve{M}-1)$. Since the elements in $\mathcal{L}^-_{np}$ are a mirrored version of the elements in $\mathcal{L}^+_{np}$, the same holds true for the set $\mathcal{L}^-_{np}$.

3. Let $l_{c1} = \breve{M}n_1 - Nm_1$ be an element of set $\mathcal{L}^+_p$ and let $l_{c2} = Nm_2 - \breve{M}n_2$ be an element of set $\mathcal{L}^-_p$. Let us assume that there exists some $(n_2, m_2)$ such that $l_{c2} = l_{c1}$. This implies that $\breve{M}n_1 - Nm_1 = Nm_2 - \breve{M}n_2$ and hence, $\frac{\breve{M}}{N} = \frac{m_1+m_2}{n_1+n_2}$. Therefore $m_2 = \breve{M} - m_1$ and $n_2 = N - n_1$. For the range of $m_1$ and $n_1$ in $\mathcal{L}^+_p$ we get $1 \leq m_2 \leq \breve{M}-1$ and $1 \leq n_2 \leq N-1$ which is within the required range for $\mathcal{L}^-_p$, implying that an element of set $\mathcal{L}^+_p$ is also an element of set $\mathcal{L}^-_p$. Hence the union of the two sets $\mathcal{L}_p$ does not contain any new unique difference than the ones available in set $\mathcal{L}^+_p$ (or $\mathcal{L}^-_p$), thus proving the claim.

4. Let $l_{c1} = \tilde{M}n_1 - Nm_1$ be an element of set $\mathcal{L}^+_{np}$ and let $l_{c2} = Nm_2 - \tilde{M}n_2$ be an element of set $\mathcal{L}^-_{np}$.
Let us assume that there exists $(n_2, m_2)$ such that $l_{c2} = l_{c1}$. This implies, $\frac{\tilde{M}}{N} = \frac{m_1 + m_2}{n_1 + n_2}$. Substituting the range of $(m_1, n_1)$ leads to the same equations obtained for the proof of Proposition I-4 and it has been shown that $m_2$ does not satisfy this range. This implies that an element in the set $\mathcal{L}^+_{np}$ is not an element in the set $\mathcal{L}^-_{np}$. Therefore, union of the two sets contain twice the number of unique differences than that contained in $\mathcal{L}^+_{np}$ (or $\mathcal{L}^-_{np}$), proving the claim.

5. Let $l_c = \tilde{M}n - Nm$ be an element of set $\mathcal{L}^+_C$. When $m=0$ we have $l_c = \tilde{M}n \in \mathcal{L}^+_{S\tilde{M}}$. When $n=0$ we have $l_c = -Nm \in \mathcal{L}^-_{SN}$. For $n \in [1, N-1]$ and $m \in [1, \tilde{M}-1]$ where $\tilde{M} < M$ we have the set $\mathcal{L}^+_p$. When $m = \tilde{M}$ and $n \in [1, N-1]$ we have $l_c = \tilde{M}n - Nm = \tilde{M}(n-N)$. The range of $(n-N)$ is $-(N-1) \leq (n-N) \leq -1$, implying that $l_c \in \mathcal{L}^-_{S\tilde{M}} - \{0\}$. Finally, for $m \in [\tilde{M}+1, M-1]$ and $n \in [1, N-1]$ we have the set $\mathcal{L}^+_{np}$ (refer Fig. 3(c)). Similarly, the relation for the set $\mathcal{L}^-_C$ can be shown to be true.

6. From Proposition II-5, we can write $\mathcal{L}_C = \mathcal{L}^+_C \cup \mathcal{L}^-_C = \mathcal{L}_S \cup \mathcal{L}_p \cup \mathcal{L}_{np}$. Since $\mathcal{L}_C$ is the union of three disjoint sets, the number of unique differences is given by the sum of the number of unique differences in each of the three sets i.e. $\{2(M+N-1)-1\} + \{(\tilde{M}-1)(N-1)\} + \{2(N-1)(M-\tilde{M}-1)\} = N(2M-1) - \tilde{M}(N-1)$ and is valid for $2 \leq p \leq M-1$. Since $\mathcal{L}_{np}$ and $\mathcal{L}_p$ are empty for $p=1$ and $p=M$ respectively, the general expression needs to appropriately include the unique differences from these sets and is given by (7).

## APPENDIX C

### PROOF OF PROPOSITION III

Proposition III-1 to III-4 are straightforward and can be proved along similar lines as Proposition 1-(a), (b) in [18]. The proof for Proposition III-5 is given below:

*Proof*: From Proposition III-3 and III-4, it follows that set $\mathcal{L}_C$ has continuous integers in the range $-MN + \tilde{M}(N-1) + 1 \leq l_c \leq MN - \tilde{M}(N-1) - 1$.
Given $l_c = \tilde{M}n - Nm$ in set $\mathcal{L}^+_C$, let us assume that $\pm(MN - \tilde{M}(N-1))$ is not a hole and exists in set $\mathcal{L}^+_C$.

This implies that $\tilde{M}n - Nm = \pm(MN - \tilde{M}(N-1))$, which can be rearranged to give $\frac{\tilde{M}}{N} = \frac{m \pm M \mp \tilde{M}}{n \mp 1}$. This does not hold true since the ratio $\frac{m + M - \tilde{M}}{n-1}$ has a denominator less than N, and the ratio $\frac{m - M + \tilde{M}}{n+1}$ can generate a denominator with value N but the numerator cannot generate a value $\tilde{M}$. If $m = M-1$ then the numerator is $\tilde{M} - 1 < \tilde{M}$, while if $m=0$ then it is $\tilde{M} - M \leq 0$, and cannot generate a positive value $\tilde{M}$. Thus, proving that it is a hole.